\documentclass[10pt,conference]{IEEEtran}

\usepackage{cite}
\usepackage{amsmath,amssymb,amsfonts}
\usepackage{algorithmic}
\usepackage{graphicx}
\usepackage{textcomp}
\usepackage{xcolor}
\def\BibTeX{{\rm B\kern-.05em{\sc i\kern-.025em b}\kern-.08em
    T\kern-.1667em\lower.7ex\hbox{E}\kern-.125emX}}
    
\PassOptionsToPackage{table,xcdraw}{xcolor}

\AtBeginDocument{%
  \providecommand\BibTeX{{%
    Bib\TeX}}}

\usepackage[english]{babel}
\usepackage[utf8]{inputenc}
\usepackage{flushend}
\usepackage{balance}
\usepackage{multirow}
\usepackage{color}
\usepackage{booktabs}
\usepackage{fancybox}
\usepackage{xcolor}
\usepackage{float}
\usepackage{array,graphicx}
\usepackage[flushleft]{threeparttable}
\usepackage[most]{tcolorbox}
\usepackage{xspace}
\usepackage{listings}
\usepackage{pifont}
\usepackage{tikz}
\usetikzlibrary{shapes.geometric}
\usepackage{blindtext}
\usepackage[numbers]{natbib}
\usepackage{colortbl}
\usepackage[table,xcdraw]{xcolor}
\usepackage{makecell}
\usepackage{stfloats}
\usepackage{url}
\usepackage{comment}

\definecolor{purplish}{HTML}{D8D0E3}
\definecolor{purplishlight}{HTML}{EBE7F1}
\definecolor{purplishdark}{HTML}{eb2d2d}

\newtcolorbox[auto counter,number within=section]{rqbox}[2]{
    nameref=#1,
    title=\small{#1}, 
    enhanced,
    attach boxed title to top left={yshift=-6pt, xshift=8pt},
    boxed title style={size=small,boxsep=1pt},
    colframe=purplishdark,colback=white,colbacktitle=purplishdark,
    boxsep=2pt,left=2pt,right=2pt,top=6pt,bottom=2pt,middle=2pt
}

\begin{document}
\title{Game Elements to Engage Students Learning the Open Source Software Contribution Process}


\author{\IEEEauthorblockN{Italo Santos\IEEEauthorrefmark{1}, Katia Romero Felizardo\IEEEauthorrefmark{1}\IEEEauthorrefmark{2}, Marco A. Gerosa\IEEEauthorrefmark{1}, and Igor Steinmacher\IEEEauthorrefmark{1}}
\IEEEauthorblockA{\IEEEauthorrefmark{1}Northern Arizona University, Flagstaff, AZ, USA\\}
\IEEEauthorblockA{\IEEEauthorrefmark{2}Federal University of Technology, Paraná, PR, Brazil\\}
Email: italo\_santos@nau.edu, katiascannavino@utfpr.edu.br, \\marco.gerosa@nau.edu, igor.steinmacher@nau.edu}

\maketitle

\begin{abstract}
    Contributing to OSS projects can help students to enhance their skills and expand their professional networks.  
    However, novice contributors often feel discouraged due to various barriers. Gamification techniques hold the potential to foster engagement and facilitate the learning process. Nevertheless, it is unknown which game elements are effective in this context.
    This study explores students' perceptions of gamification elements to inform the design of a gamified learning environment.
    We surveyed 115 students and segmented the analysis from three perspectives: (1) \textit{cognitive styles}, (2) \textit{gender}, and (3) \textit{ethnicity} (Hispanic/LatinX and Non-Hispanic/LatinX).
    The results showed that Quest, Point, Stats, and Badge are favored elements, while competition and pressure-related are less preferred. Across cognitive styles (persona), gender, and ethnicity, we could not observe any statistical differences, except for Tim's GenderMag persona, which demonstrated a higher preference for storytelling. Conversely, Hispanic/LatinX participants showed a preference for the Choice element.
    These results can guide tool builders in designing effective gamified learning environments focused on the OSS contributions process.  
\end{abstract}

\textbf{\textit{keywords: gamification, game elements, human factors, cognitive styles, human-computer interaction.}}

\section{Introduction}
\label{sec:introduction}

Open Source Software (OSS) projects are not just code repositories, they are evolving ecosystems where learning and collaboration converge. OSS initiatives receive a worldwide network of contributors partly due to the growing recognition of the value that experience with OSS projects offers to professionals~\cite{gerosa2021shifting}. For students, in particular, contributing to OSS projects is an effective way to develop practical skills, gain real-life experience, and boost their employability~\cite{jergensen2011onion, goduguluri2011kommgame,silva2020google}.

Given the importance of providing students with practical skills, educators integrate OSS projects into their curriculum to expose students to authentic software development challenges~\cite{pinto2019training, pinto2017training, ferreira2018students}. However, students new to OSS, especially those from diverse backgrounds, including different cognitive styles, genders, and ethnicities, might feel demotivated from participating due to barriers that prevent contributions from newcomers~\cite{steinmacher2015social, mendez2018open, padala2020gender,trinkenreich2021women}. Engaging and motivating students to contribute to OSS projects is a challenge~\cite{PS04diniz2017using, muller2019engaging, flach2023teaching}.

A potential strategy to engage students is employing gamification, which increases motivation by integrating game elements into a system~\cite{dichev2017gamifying}. Through gamification, students are more likely to remain committed, find joy in their contributions, and stand to benefit educationally~\cite{pedreira2015gamification, dicheva2015gamification, bartel2016gamifying}. For example, Diniz et al.~\cite{PS04diniz2017using} implemented gamification strategies to increase motivation and facilitate collaboration among undergraduate students in OSS projects using GitLab. They evidenced that game elements, such as quests and points, kept students engaged and aided in orienting and motivating them to contribute to the projects. Nevertheless, no studies have examined the appropriateness of a comprehensive set of game elements designed to foster student engagement in learning the OSS contribution process.

Our study investigates students' perceptions of game elements for this purpose. We particularly focus on understanding how the perception changes across diverse cognitive styles, gender identities, and ethnic backgrounds. 

The following research questions guided our investigation:

\newcommand{\rqtextone}{Which game elements do students prefer to engage in learning the OSS contribution process?}
\newcommand{\rqone}[2][]{
    \begin{rqbox}{\textbf{Research Question 1}}{#2}
        \rqtextone
        #1
    \end{rqbox}
}

\rqone{}

\newcommand{\rqtexttwo}{How do students' preferences of game elements differ according to their cognitive style, gender, and ethnicity?}
\newcommand{\rqtwo}[2][]{
    \begin{rqbox}{\textbf{Research Question 2}}{#2}
        \rqtexttwo
        #1
    \end{rqbox}
}

\rqtwo{}

To answer our research questions, we surveyed 115 students and identified their preferences for various game elements. We designed a survey to collect responses from students majoring in computer science and related fields (e.g., applied computer science, software engineering, information systems, and computer engineering). The findings from this survey can inform future targeted experimental research, where specific game elements can be implemented and their effects directly measured. While integrating actual game elements into the open-source tools would be informative, this study primarily aims to explore the receptiveness to such mechanisms. Understanding whether and which game elements are perceived as valuable is a necessary precursor to their practical application. Gathering potential end users' perceptions before implementing a tool not only contributes to usability and satisfaction but also offers valuable insights from the user-centered design perspective. 

Our findings suggest that game elements characterized by performance-oriented features, such as Stats, Maps, Levels, Points, Progress Bars, and Badges, are preferred for student engagement. These elements not only provide informative feedback on performance but also contribute to the development of a sense of progression and achievement. With a few exceptions, we did not find much difference across demographics. We expect that our insights on students' preferences for game elements will help the research community create gamified learning environments focused on the OSS contribution process. 

\section{Background and Related Work}
\label{sec:backrelated}

According to Ghai and Tandon~\cite{ghai2023integrating}, gamification has become a popular topic among educators due to recent technological advancements and a growing interest in innovative educational methods. Gamification gained popularity due to its ability to engage and inspire students, helping them achieve a state of flow throughout the learning process~\cite{huotari2017definition}. Dichev and Dicheva~\cite{dichev2017gamifying} describe gamification as a strategy to enhance student motivation and engagement in educational settings by integrating game elements. Gamification leverages game elements' inherent motivational appeal and immersive qualities to encourage student participation in the learning process. Deterding et al.~\cite{deterding2011gamification} suggest that gamification simplifies complex material, making learning more approachable and facilitating a deeper understanding of the subject matter. They consider that games can transform educational experiences, making them more interactive, enjoyable, and effective for students.

Zhao et al.~\cite{zhao152023motivating} focused on encouraging contributions to OSS through a network algorithm called OpenRank and a monthly updated contribution leaderboard. The results show that gamified leaderboards can impact OSS developers towards better collaboration and a more harmonious community atmosphere with social connections. Kapp~\cite{kapp2012gamification} salients the transformative power of gamification, observing that, when implemented effectively, it can captivate and educate students. Huotari and Hamari~\cite{huotari2012defining} further assert that gamification, when utilized as an educational tool, extends beyond merely expanding student knowledge; it also enhances their collaborative and communicative skills. In education, gamification introduces game-based rules, immersive player experiences, and cultural contexts, which collectively influence and modify student behavior~\cite{lee2011gamification}. 

Recent studies have explored the application of gamified environments to software engineering education. Su~\cite{su2016effects} created a gamified framework to examine the impact of gamification on teaching software engineering. The findings indicated that students were more motivated by gamified teaching methods, which enhanced academic outcomes. Similarly, Sheth et al.~\cite{sheth2012increasing} demonstrated that introducing gamification to a software engineering course could boost student engagement in various aspects of the development process, including documentation, bug reporting, and testing activities. 

Diniz et al.~\cite{PS04diniz2017using} implemented gamification strategies to motivate and support student collaboration in OSS projects using GitLab. Their findings highlighted that game elements, such as quests and points, kept students engaged and aided in orienting and motivating them to contribute to OSS projects. Toscani et al.~\cite{PS17toscani2018gamification} demonstrated that gamification can effectively engage a diverse range of newcomers. Moreover, gamification has been successfully applied to various domains within software engineering education, including agile process training~\cite{prause2012field}, mutant testing~\cite{rojas2016teaching}, and the learning of design patterns~\cite{bartel2016gamifying}, Graphical User Interface (GUI) testing~\cite{garaccione2022gerry}, cybersecurity training~\cite{decusatis2022gamification}, exploratory testing~\cite{ozturk2022gamification}, and design patterns~\cite{bartel2016gamifying}, demonstrating its wide-ranging applicability and potential to enrich the educational experience in this field.

Our study enhances the existing body of research by exploring student perceptions regarding implementing game elements in a gamified learning environment specifically designed to educate about the OSS contribution process. We uncover student preferences for specific game elements to present a nuanced understanding of how various game elements can be considered among student profiles, thereby informing the development of more inclusive and effective gamified educational tools.

\section{Research Method}
\label{sec:methodology}

This study investigates students' perceptions of game elements in the context of learning the OSS contribution process. In this section, we detail our method. 


\subsection{Survey design}

Our survey consisted of four main sections: (Part I) scenario explanation; (Part II) game elements; (Part III) GenderMag facets; and (Part IV) demographics. We used open, multiple-choice, and 5-Likert scale questions (from 1-- strongly disagree to 5--strongly agree). We implemented the survey in Qualtrics\footnote{www.qualtrics.com}.

Below, we briefly explain the survey sections:

\textbf{Part I -- Scenario Explanation:} Respondents were directed to a consent page that explains the study's objectives, the rules regarding confidentiality, the estimated time commitment required to complete the survey, and contact information for the researchers. After that, to guide participants in providing relevant and context-aware responses, we offer a scenario to consider while answering questions related to game elements in the survey: \textit{You are a student in a university class using an interactive environment to learn how to contribute to Open Source Software (OSS) projects. You must accomplish tasks such as exploring projects, collaborating with fellow learners, reviewing pull requests, making non-code contributions (e.g., documentation, translation, etc.), and establishing connections with the community.}

\textbf{Part II -- Game Elements:} This section was designed to collect information on the game elements students perceive as engaging within a gamified learning environment focused on teaching the process of contributing to OSS. We used Toda et al.~\cite{toda2019analysing} taxonomy as our framework, which focuses on gamification strategies in educational contexts. The game elements included in our survey were selected from the taxonomy during design sessions, which focused on representing the game elements in the context of the OSS contribution process. During these sessions, we considered the trade-offs of including additional dimensions and implementations with the need to keep the instrument short. Toda et al.'s taxonomy is structured into five dimensions~\cite{toda2019analysing}, as outlined below:

\begin{itemize}
    \item \textbf{Performance} -- This dimension is related to responses from the environment that serve as feedback mechanisms for the learner, including elements such as points, progress bars, levels, stats, and acknowledgments, i.e., badges.
    
    \item \textbf{Ecological} -- The elements of this dimension, including choice, economy, and time pressure, act as properties of the environment to engage learners to follow a desired behavior.

    \item \textbf{Social} -- Concerning this dimension, the game elements involve interactions among learners within the environment, including cooperation, reputation, social pressure, and competition, i.e., leaderboards, private rank, and competitive tasks.

    \item \textbf{Personal} -- The elements of this dimension focus on the individual learner's engagement with the environment, featuring aspects such as objectives (e.g., quests), puzzles, novelty, and renovation.

    \item \textbf{Fictional} -- Regarding this dimension, the elements connect the user to the environment through user-oriented narratives and environment-oriented storytelling, intertwining their experience with the context.
\end{itemize}

Table~\ref{tab:surveyquestions} describes the game elements and the questions used in the survey.

Still, for each game element, we included an optional open-ended question on the survey --- \textit{Please share any additional thoughts or details to justify your rating.} --- This was done to gather insights into why participants gave specific ratings for each game element. Furthermore, we asked participants to rank the top three most relevant they would like to see while using a gamified learning environment.

\begin{table*}[!ht]
\centering
\caption{Game elements and their corresponding survey questions}
\label{tab:surveyquestions}

\begin{tabular}{l|l|l}
\hline
\begin{tabular}[c]{@{}l@{}}Taxonomy \\ dimensions\end{tabular} & \begin{tabular}[c]{@{}l@{}}Game \\ elements\end{tabular} & Survey questions \\ \hline \hline
\multirow{6}{*}{\begin{tabular}[c]{@{}l@{}}Performance\end{tabular}} & Point & \begin{tabular}[c]{@{}l@{}}After completing tasks, you receive experience points proportionally to your performance. You will likely come back \\ because you like gaining experience points.\end{tabular} \\ \cline{2-3} 
 & Level & \begin{tabular}[c]{@{}l@{}}As you complete tasks, you gain levels, and increasing your level unlocks new and more complex tasks (e.g., you can \\ only submit a contribution if you are at level “5”). You will likely come back because you like increasing your level.\end{tabular} \\ \cline{2-3} 
 & Badges & \begin{tabular}[c]{@{}l@{}}You receive a badge as an award for successfully meeting specific requirements (e.g., a ``Flash'' badge when you finish \\ tasks before their time limit, etc.). You will likely come back because you like conquering more badges.\end{tabular} \\ \cline{2-3} 
 & Progress Bar & \begin{tabular}[c]{@{}l@{}}You can see a progress bar with the accumulated experience points and the remaining points required to attain the \\ next level. You will likely come back because you like visualizing how many tasks you still have to complete to keep \\ progressing.\end{tabular} \\ \cline{2-3} 
 & Map & \begin{tabular}[c]{@{}l@{}}As you advance in your learning journey, you see a visual map of your learning progress, including completed and \\ remaining tasks you must accomplish. You will likely come back because you like visualizing an overall map with all \\ the tasks you have (and have not) completed.\end{tabular} \\ \cline{2-3} 
 & Stats & \begin{tabular}[c]{@{}l@{}}When you interact with the learning environment, you can visualize your learning results and other relevant \\ information (e.g., current level, experience points (XP), tasks completed, badges, etc.). You will likely come back \\ because you like visualizing your overall stats.\end{tabular} \\ \hline \hline
 
\multirow{3}{*}{Ecological} & Choice & \begin{tabular}[c]{@{}l@{}}You can choose the tasks you want to work on in the learning environment, such as editing a file or reviewing a \\peer's contribution. You will likely come back because you like choosing a task to work on.\end{tabular} \\ \cline{2-3} 
 & Economy & \begin{tabular}[c]{@{}l@{}}You can do extra tasks (e.g., quizzes) to accumulate special items (e.g., gems), and you can trade the special items \\ received for advantages, such as help from the instructor/assistant to complete a task. You will likely come back \\ because you like trading items for advantages.\end{tabular} \\ \cline{2-3} 
 & \begin{tabular}[c]{@{}l@{}}Time \\ Pressure\end{tabular} & \begin{tabular}[c]{@{}l@{}}You have to complete a task (e.g., submit the assignment) in the given timeframe. Once the task has been opened, \\ the clock starts counting backward. If you complete the task before the specific time, you will be rewarded. You will \\ likely come back because you like doing tasks with time pressure.\end{tabular} \\ \hline \hline
 
\multirow{5}{*}{Social} & Cooperation & \begin{tabular}[c]{@{}l@{}}You can perform collaborative tasks, i.e., tasks involving other peers. You will likely come back because you like\\ working on collaborative tasks.\end{tabular} \\ \cline{2-3} 
 & \begin{tabular}[c]{@{}l@{}}Social \\ Pressure\end{tabular} & \begin{tabular}[c]{@{}l@{}}The learning environment exposes your task results for other learners to peer-review. You will likely come back\\ because you like to show your work to other learners.\end{tabular} \\ \cline{2-3} 
 & \begin{tabular}[c]{@{}l@{}}Public \\ Leaderboard\end{tabular} & \begin{tabular}[c]{@{}l@{}}In the learning environment, you see a public leaderboard and can compare your performance to others (e.g., a public \\leaderboard with a ranking of all learners in the environment). You will likely return because you like improving\\ your performance based on the public leaderboard.\end{tabular} \\ \cline{2-3} 
 & Private Rank & \begin{tabular}[c]{@{}l@{}}You see your position in a private rank compared to your peers (e.g., three learners have more points than you; you\\ are 5 points behind the 3rd position). You will likely return because you like improving your position in the\\ ranking.\end{tabular} \\ \cline{2-3} 
 & \begin{tabular}[c]{@{}l@{}}Competitive \\ Task\end{tabular} & \begin{tabular}[c]{@{}l@{}}You engage in tasks competitively with other learners to determine who achieves the best results (e.g., you have to\\ finish a before all other learners). You will likely come back because you like completing tasks and competing \\against other learners.\end{tabular} \\ \hline \hline
 
\multirow{4}{*}{Personal} & Quest & \begin{tabular}[c]{@{}l@{}}The environment offers quests to guide your learning journey. Each quest involves tasks such as editing a file. \\ You will likely come back because you like accomplishing quests.\end{tabular} \\ \cline{2-3} 
 & Puzzle & \begin{tabular}[c]{@{}l@{}}You can perform challenging tasks related to the content learned (e.g., contribute to a project outside of the learning\\ environment) to measure your learning. You will likely come back because you like being challenged with tasks \\related to the content learned.\end{tabular} \\ \cline{2-3} 
 & Renovation & \begin{tabular}[c]{@{}l@{}}You can redo a task that you fail, such as committing your changes with clearer commit messages, to increase your \\ score or experience points. You will likely come back because you like redoing a failed task.\end{tabular} \\ \cline{2-3} 
 & Novelty & \begin{tabular}[c]{@{}l@{}}The learning environment offers new content, such as unlocking new tasks, including novelty, as you advance in \\ your learning process. You will likely come back because you like having new content.\end{tabular} \\ \hline \hline
 
\multirow{2}{*}{Fictional} & Narrative & \begin{tabular}[c]{@{}l@{}}The learning environment presents a dystopic narrative (e.g., about an individual who needs to learn about the OSS \\ process to improve the quality of life in society). You must interact with the environment to advance in the storyline. \\ You will likely come back because you like having a narrative.\end{tabular} \\ \cline{2-3} 
 & Storytelling & \begin{tabular}[c]{@{}l@{}}The learning environment shows the storytelling using images and text related to the narrative. You will likely come\\ back because you like storytelling.\end{tabular} \\ \hline \hline
\end{tabular}
\end{table*}

\textbf{Part III -- GenderMag Facets:} In this section, participants were asked how they behave when they approach unknown technology. The questions were used to evaluate participants' GenderMag~\cite{burnett2016gendermag} facets. GenderMag is a systematic inspection method that captures individual differences in how people solve problems and use software features. GenderMag is based on research showing that individual differences in cognitive styles (facets) cluster by gender~\cite{murphy2024gendermag, santos2023designing, hilderbrand2020engineering}. The method encapsulates the facets into personas (Abi and Tim) related to (i) Motivation: Abis are motivated to use technology for what they can accomplish with it, whereas Tims are often motivated by their enjoyment of technology per se; (ii) Information processing styles: Abis process new information comprehensively--gathering fairly complete information before proceeding--but Tims use selective information processing--following the first promising information, then backtracking if needed; (iii) Computer self-efficacy: relates to a person’s confidence about succeeding at a specific task, which influences their use of cognitive strategies, persistence, and strategies for coping with obstacles. Abis have lower computer self-efficacy as compared to their peers; (iv) Risk aversion: Abis are risk-averse when trying out new features as compared to Tims, which impacts their decisions about which feature sets to use; and (v) Learning: by Process vs. by Tinkering: Abis prefer process-oriented learning, whereas Tims, like to experiment (``tinker'') with software features new to them. Each cognitive style has advantages, but it is also at a disadvantage when not supported by the software. 

In summary, Abi and Tim are in the opposite spectrum of facet values, with the Abi persona aligned with facet values that women tend to favor and Tim embodying facet values typically favored by men. We selected five questions from the complete GenderMag to make it more manageable by shortening the number of questions. We selected these questions by applying the variance inflation factor (VIF) on the data of a previous study~\cite{santos2023designing} that collected participants' information using the complete GenderMag questionnaire to reduce the Multicollinearity~\cite{james2013introduction}. Based on participants' answers, we could characterize participants' cognitive styles according to the facets of each GenderMag persona (e.g., Tim or Abi).

\textbf{Part IV -- Demographics:} Participants were asked to provide information about age, the country where they live, gender, ethnicity, the year in college they are enrolled in, if they have taken a course about OSS, and years of experience working in software development, games, and gamified learning environments.

The survey was anonymous. The order of the game elements that appear in the response options was randomized to mitigate response order effects~\cite{lavrakas2008encyclopedia}. All open questions were optional to increase the response rate by making respondents more comfortable~\cite{punter2003conducting}. Furthermore, an attention check question was also included to ensure that participants read instructions carefully~\cite{danilova2021you}: ``This is an attention check; please mark the option Strongly Disagree.'' 

\subsection{Pilot}

After we designed and proofread the survey, we tested it in multiple browsers and devices. We invited five participants to pilot the survey to ensure the survey language was appropriate to our target population, collect feedback, and measure the survey response time. We had three pilot sessions with undergraduate computer science students that fit the respondent profile. In those sessions, we validated if the students could understand the questions well. The participants suggested some language improvements to make the survey more straightforward. One example was the initially proposed question for the level element: \textit{You will likely come back because you like gain levels}. During the pilot, the participants mentioned that the word \textit{increase} is more common in a game context than \textit{gain}. After the improvements made in the survey from this first interaction, we conducted the pilot study with two open-source specialists. We received more feedback about the content presented in the survey. Based on the feedback, we made some changes to improve our questions, dropping some questions to reduce the survey length and including some new questions to validate that the participants are part of our target population.

\subsection{Recruitment}

The respondents were recruited using Prolific\footnote{www.prolific.co}, a crowdsourcing platform where researchers can channel their surveys~\cite{russo2022recruiting}. Prolific keeps the demographic information of all its users to facilitate the pre-selection process. We could choose some pre-screening characteristics to narrow down the population to our relevant populations. We focused on undergrad students taking any CS-related major. We created four recruitment surveys on Prolific, with the following extra criteria based on our population of interest: (i) Men and Hispanic/LatinX; (ii) Non-Men and Hispanic/LatinX; (iii) Men and Non-Hispanic/LatinX; (iv) Non-Men and Non-Hispanic/LatinX. We focused on Hispanic/LatinX population because it is chronically underrepresented in computer science and OSS~\cite{villegas2024us}. Our goal was to explore how cultural differences might correlate with preferences for certain game elements. We aimed for a balanced distribution to compare responses across different groups.

We included screening questions at the beginning of our survey to check that the respondents still hold the same demographic information as when they filled out their information to Prolific. Therefore, we asked the same questions asked by the prolific platform that we used to filter the population. After answering the pre-screening validation section, respondents could access the complete survey (in case they were still part of the group of interest). 

The survey was conducted from February 15 to March 10, 2024, and participants were paid \$3.0 through the Prolific platform after completing the survey. We received 158 responses that were not blank. Following a thorough review, detailed in the subsequent subsection, we identified 115 valid responses for our analysis.

\subsection{Filtering}

We carefully reviewed and filtered our data to consider only valid responses. We dropped answers that failed the attention check question (five cases) and checked for answers with the same choice for all Likert scale questions (0 removed). We removed participants who did not complete the survey (16 cases) and those who failed the prescreening questions (22 cases). We then analyzed the time to complete the survey to remove lower outliers (0 removed). After this filtering process, we end up with 115 valid responses for analysis in our dataset.

\subsection{Data analysis}

\textit{1) Likert-scale items:} We asked the level of agreement of participants using Likert-scale items (ranging from 1 representing strongly disagree to 5 representing strongly agree) for a set of game elements. Then, participants ranked the top three game elements they would like to see in a learning environment proposed. We used descriptive statistics and visually analyzed the data. We segmented our sample based on \textit{cognitive styles} (persona: Abi and Tim), \textit{gender} (man and minorities --- women and non-binary), and \textit{ethnicity} (Hispanic/LatinX and Non- Hispanic/LatinX) and checked the odds ratios to evaluate the interest of subgroups and Mann-Whitney U test to analyze the differences in game element preferences across them.

\textit{2) Open questions:} We used open questions to ask for more information about participants' perceptions of the game elements. We qualitatively analyzed participants' comments following open coding procedures~\cite{strauss1998basics}. The process was conducted using continuous comparison and discussion until reaching a consensus. Two researchers jointly analyzed the sets of answers to establish common ground, discussing the applied codes and disagreements until reaching a consensus. Finally, a third researcher inspected the classification.

The replication package, with the anonymized dataset, instruments, and scripts, is available for public access~\footnote{\url{https://figshare.com/s/5b5a16dcb17d16781bae}}.

\section{Participant Demographics}
\label{sec:demographics}

The demographics of our participants are summarized in Table~\ref{tab:demographics}. Our demographics show a gender distribution with the majority identifying as men (73\%), followed by women (31\%), and a smaller representation of non-binary respondents (5\%). Concerning ethnic identity among our respondents, we have 49\% of respondents as Hispanic/LatinX, with the remaining 51\% not identifying as Hispanic/LatinX. As expected, the age distribution of our survey participants indicates a young demographic, with 48\% being 24 or below, 39\% between the ages of 25 and 34, and smaller percentages distributed among older age groups. In terms of academic major, the largest group of respondents were studying computer science (43\%), followed by other computing-related majors/programs (17\%), information technology (15\%), and software engineering (13\%), among others. Looking at their academic journey, 37\% were in their $4^{th}$ or $5^{th}$ year, indicating a senior level of study among the respondents. The survey also touched upon the GenderMag personas, where 54\% aligned with the persona ``Abi'' and 46\% with ``Tim'', indicating a diverse range of user perspectives.

\begin{table}[!ht]
\centering
\caption{Personal characteristics of respondents (N=115)}
\label{tab:demographics}
\begin{tabular}{lcc}
\hline
\multicolumn{1}{c|}{\textbf{Demographics}} & \multicolumn{1}{c|}{\#} & \% \\ \hline

\hline

\multicolumn{3}{c}{\textbf{Gender}} \\ \hline \hline
\multicolumn{1}{l|}{Man} & \multicolumn{1}{c|}{\textbf{73}} & \textbf{64\%} \\ \hline
\multicolumn{1}{l|}{Woman} & \multicolumn{1}{c|}{\textbf{36}} & \textbf{31\%} \\ \hline
\multicolumn{1}{l|}{Non-binary} & \multicolumn{1}{c|}{\textbf{6}} & \textbf{5\%} \\ \hline \hline

\multicolumn{3}{c}{\textbf{Ethnicity}} \\ \hline \hline
\multicolumn{1}{l|}{Hispanic/LatinX} & \multicolumn{1}{c|}{\textbf{56}} & \textbf{49\%} \\ \hline
\multicolumn{1}{l|}{Not Hispanic/LatinX} & \multicolumn{1}{c|}{\textbf{59}} & \textbf{51\%} \\ \hline \hline

\multicolumn{3}{c}{\textbf{Age}} \\ \hline \hline
\multicolumn{1}{l|}{24 or below} & \multicolumn{1}{c|}{\textbf{56}} & \textbf{48\%} \\ \hline
\multicolumn{1}{l|}{25 to 34} & \multicolumn{1}{c|}{\textbf{45}} & \textbf{39\%} \\ \hline
\multicolumn{1}{l|}{35 to 44} & \multicolumn{1}{c|}{\textbf{10}} & \textbf{9\%} \\ \hline
\multicolumn{1}{l|}{45 to 54} & \multicolumn{1}{c|}{\textbf{3}} & \textbf{3\%} \\ \hline
\multicolumn{1}{l|}{55 to 64} & \multicolumn{1}{c|}{\textbf{1}} & \textbf{1\%} \\ \hline \hline

\multicolumn{3}{c}{\textbf{Year of Study}} \\ \hline \hline
\multicolumn{1}{l|}{1st year (freshman)} & \multicolumn{1}{c|}{\textbf{18}} & \textbf{16\%} \\ \hline
\multicolumn{1}{l|}{2nd year (sophomore)} & \multicolumn{1}{c|}{\textbf{22}} & \textbf{19\%} \\ \hline
\multicolumn{1}{l|}{3rd year (junior)} & \multicolumn{1}{c|}{\textbf{27}} & \textbf{23\%} \\ \hline
\multicolumn{1}{l|}{4th or 5th year (senior)} & \multicolumn{1}{c|}{\textbf{42}} & \textbf{37\%} \\ \hline
\multicolumn{1}{l|}{Graduate studies} & \multicolumn{1}{c|}{\textbf{6}} & \textbf{5\%} \\ \hline \hline
\multicolumn{3}{c}{\textbf{GenderMag Persona}} \\ \hline \hline
\multicolumn{1}{l|}{Abi} & \multicolumn{1}{c|}{\textbf{62}} & \textbf{54\%} \\ \hline
\multicolumn{1}{l|}{Tim} & \multicolumn{1}{c|}{\textbf{53}} & \textbf{46\%} \\ \hline \hline

\end{tabular}
\end{table}

\section{Results}
\label{sec:results}

\subsection{Students' perception about game elements}

To answer RQ1: \textit{\rqtextone}, we asked participants Likert-scale items to rate their preference. Participants also ranked their top three game elements and expressed their rationale through open questions.

\begin{figure*}[!ht]
    \centering
    \includegraphics[width=1.0\textwidth]{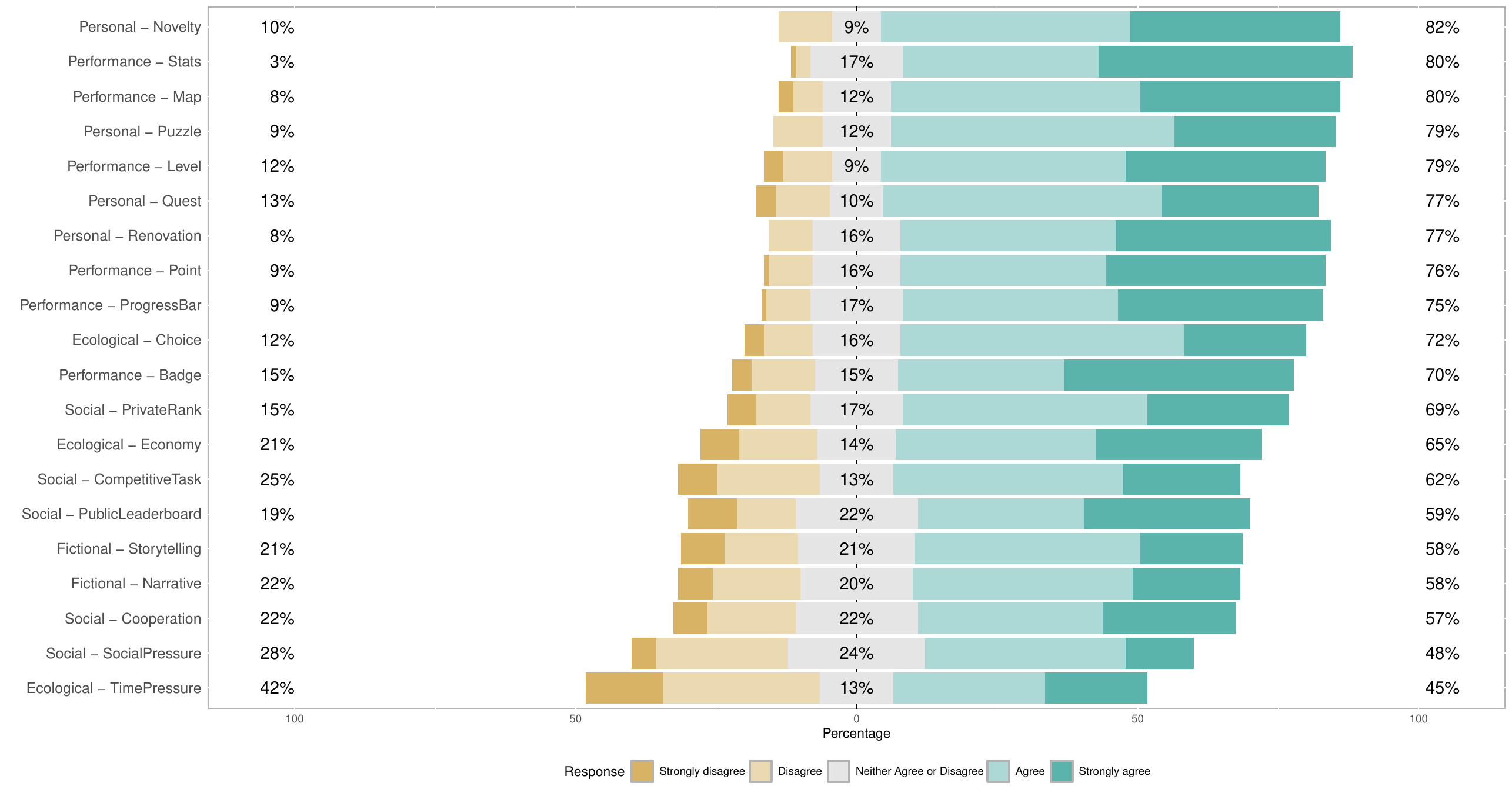}
    \caption{Responses to the 5-point Likert-scale items for the game elements. The left hand (yellow) shows levels of disagreement, the middle (grey) shows neutral, and the right (green) shows levels of agreement.}
    \label{fig:likertGameElements}
\end{figure*}

\textbf{Students preferences for game elements.} Figure~\ref{fig:likertGameElements} shows the answers to the Likert-scale items. Our findings indicate that the respondents agree (or strongly agree) that \textit{Performance dimension} elements contribute to their engagement. Those elements refer to stats and maps (80\%), levels (79\%), points (76\%), progress bars (75\%), and badges (70\%). Participants underscore the significance of such feedback in enhancing task completion, as stated by P13: ``\textit{Gaining experience to gain levels makes me feel like I'm working towards something important.}''. Toda et al.~\cite{toda2019analysing} emphasize that the \textit{performance dimension} is critical in providing feedback for students.

Additionally, respondents agree that \textit{Personal dimension} elements contribute to their engagement with the environment. Those elements include novelty (82\%), puzzles (79\%), quests (77\%), and renovation (77\%), with P41 explaining that: ``\textit{I appreciate having new content introduced in the learning environment because it keeps me engaged and motivated by offering fresh challenges and opportunities for growth.}''. 

Within the \textit{Ecological dimension}, 72\% of respondents agree that the choice element allows students to tailor their experience by optimizing their interaction with the environment based on their preferences and available time. For instance, P10 highlighted the practicality of this element: ``\textit{choosing tasks is useful, so you can measure the cost-benefit of doing them based on the time available.}''. Furthermore, while 65\% of participants agree (or strongly agree) that they enjoy trading items for advantages, 21\% disagree (or strongly disagree). P13 expressed enjoyment and a sense of achievement from collecting items. In contrast, P31 raised concerns about fairness, suggesting, ``\textit{I agree the use of special items to trade for special advantages outside a learning environment, but these advantages should be available to everyone in a learning environment and not only to those who can pay for it}''.

Relative to the \textit{Social dimension} and their elements, while most participants perceive the value of Private Rank, Competitive Task, and Public Leaderboard elements, with agreement rates at 69\%, 62\%, and 59\%, respectively, a minority expressed reservations, with 15\%, 25\%, and 19\% disagreeing. Student feedback reveals diverse perspectives. On private ranks, one student (P47) states: ``\textit{if I know how I am doing compared to others privately it will inspire me to do better}'' suggesting a motivational benefit. In contrast, P44 resists the competitive approach, stating, ``\textit{I don't view education as a competition.}''. Regarding competitive tasks, P88 comprehends that ``\textit{The competition aspect would make me get more involved.}'' while P78 express discomfort, noting, ``\textit{I would not like to participate in competitions. They make me nervous, and the opponent is always overly confident.}''. Public leaderboards also draw mixed feelings. P42 notes motivation in competition, saying, ``\textit{This creates competition, which is often a great motivator.}'' Meanwhile, P50 articulates concern about the implications of comparing grades publicly, manifesting: ``\textit{I would likely be a bit competitive, but I don't like the idea of a public leaderboard in a class if it's comparing grades.}''.

The \textit{Fictional dimension}, encompassing Storytelling and Narrative, aims to bridge the user with the environment, enhancing the learning experience by embedding it within a broader context. However, opinions on these elements are divided, with roughly 21\% and 22\% of respondents showing ambivalence despite an average agreement rate of 58\%. For Storytelling, one participant (P44) admits that integrating related images and text makes learning more memorable and engaging. In contrast, another participant (P77) thinks such elements can detract from the learning focus. As for Narrative, some see it as a tool to forge meaningful connections, with P42 noting its potential to make content more relatable and encourage more profound commitment. However, P23 criticizes it for feeling juvenile and counterproductive, suggesting narratives be more aligned with real-world scenarios and professional language, thereby preserving the learning's relevance and respectability. 

Overall, the elements Time Pressure, Social Pressure, and Cooperation received the highest disagreement rate from participants (42\%, 28\%, and 22\%, respectively). For instance, P52 cited stress induced by time constraints, while P104 voiced discomfort with their work being visible to peers, leading to overthinking. P74's reluctance towards cooperative tasks was influenced by past negative experiences, often with unequal workload distribution.

\begin{figure*}[!ht]
    \centering
    \includegraphics[width=1.0\textwidth]{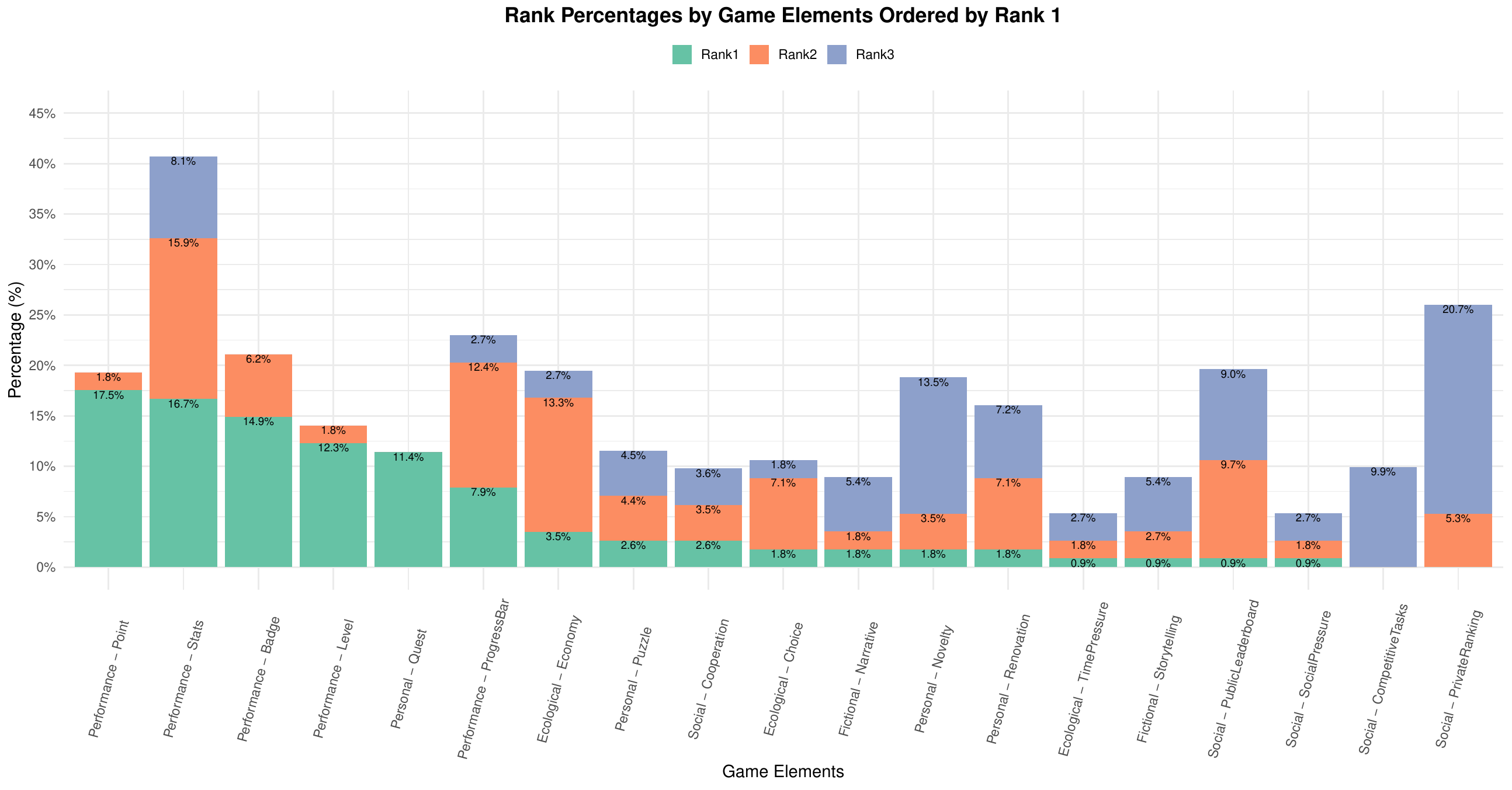}
    \caption{Responses to the top 3 ranked game elements ordered by rank 1. Each bar represents the percentage of respondents who ranked a particular game element as their top three preferences.}
    \label{fig:toprank}
\end{figure*}

\textbf{Ranked game elements.} Participants identified the top three elements they preferred for inclusion in the gamified learning environment. The results are visually represented in a stacked bar plot, as shown in Figure~\ref{fig:toprank}. This plot categorizes the game elements by their rankings---from 1 to 3, with 1 being the highest priority.

Reflecting on the elements that received the highest rankings in Rank 1, it is clear that those from the performance dimension---namely Point, Stats, Badge, and Level---stand out, garnering approximately 17.5\%, 16.7\%, 14.9\%, and 12.3\% of the first-place votes, respectively. Stats, in particular, demonstrate a strong preference among participants, achieving significant percentages across the top three ranks, marking an inclination towards performance-related game elements.

Remarkably, the quest element, capturing 11.4\% of the top-rank votes, stands out as the singular element consistently preferred as the top choice, suggesting distinctive initial favoritism among participants. Meanwhile, ProgressBar and Economy showcase a uniform distribution of preferences across all three rankings (23\% and 19.5\%, respectively). 

Other elements, including Puzzle (11.5\%), Collaboration (9.7\%), Renovation (16.1\%), and Novelty (18.8\%), exhibit moderate levels of preference, with their distribution across the three rankings indicating diverse views on their significance and attractiveness. This observation is intriguing given that, in Figure~\ref{fig:likertGameElements}, Novelty emerged as the element with the highest level of agreement.

Additionally, it is worth noting that the Map element, despite receiving a higher rating in Figure~\ref{fig:likertGameElements}, was absent from participants' rankings. The omission may suggest that participants preferred the ProgressBar over the Map as a progression mechanism.


\textbf{Student's perceptions.} As previously mentioned, in our qualitative analysis, we adopted open coding to identify related content across participants' responses, with \textit{Engagement} emerging universally across all game elements. Another noteworthy category we unveiled, \textit{Fun}, was referenced for nearly all elements, excluding Progress Bar, Public Leaderboard, Private Rank, Social Pressure, and Renovation. These exceptions suggest that aspects such as the visual progress indicators from the progress bar, the competition, and the pressure within the gamified setting are not universally perceived as enjoyable. The finding was surprising, especially since Renovation was highly rated on our scale (Figure~\ref{fig:likertGameElements}). However, the discrepancy could be clarified by participant P67's remark: ``This is where it would start to get bothersome. I know I need help learning. Just not sure this would work for my type of personality.'' The comment suggests that while the concept of Renovation is appealing, its application may not align with everyone's learning preferences or personality types, reflecting the nuanced impact of gamified elements on individual user experience.

\textit{Autonomy} emerged only when participants commented about the Choice element, highlighting the significance of providing students with decision-making. P44 elaborated on its value, stating, ``\textit{ I like having the ability to focus on the things I enjoy doing, or feel as though I do a good job at.}''. The affirmation highlights how autonomy through choice enhances user engagement by allowing personal interests and strengths to guide their actions.

We also identified that participants like the \textit{Sense of Progress} associated with game elements. This highlights the role of gamification in advancing students' perceptions of progression. However, this perception is less associated with Public Leaderboard and Puzzle elements. It is inferred that the Leaderboard may predominantly evoke a sense of competition. At the same time, puzzles might be viewed more as a mechanism for assessing learning, as indicated by P38: ``\textit{It would feel REALLY good to get real-world experience and challenge myself.}''. Additionally, alongside puzzles, other elements were also recognized by participants for their role in enhancing learning, including Novelty, Renovation, Narrative, and Storytelling, as outlined by P54: ``\textit{I think this helps me learn, to receive feedback and improve on a specific task.}''.

Elements centered around competition (such as Public Leaderboard, Private Rank, and Competitive Task) and those focusing on Cooperation and Social Pressure demonstrate mixed reactions among participants. While some exhibit an apparent fondness for the competitive aspects, others show a strong dislike due to the potential stress or discomfort they could cause. Social Pressure arises as highly polarizing, with many respondents rejecting it, primarily because of its association with peer evaluation.

\rqone[
    \tcblower
    \textbf{Answer:} Elements such as Quest, Point, Stats, and Badge emerged as clear favorites. In contrast, elements associated with competition and pressure were viewed less favorably.
]{}

\begin{table}[!ht]
\centering
\caption{Odds ratios per characteristics of respondents}
\label{tab:oddsratio}
\begin{tabular}{llll}
\hline \hline
\rowcolor[HTML]{FFFFFF} 
\multicolumn{1}{l|}{\cellcolor[HTML]{FFFFFF}\textbf{}} & \multicolumn{1}{l|}{\cellcolor[HTML]{FFFFFF}\textbf{\begin{tabular}[c]{@{}l@{}}Abi vs.\\  Tim\end{tabular}}} & \multicolumn{1}{l|}{\cellcolor[HTML]{FFFFFF}\textbf{\begin{tabular}[c]{@{}l@{}}Men vs.\\  Minorities\end{tabular}}} & \multicolumn{1}{l}{\cellcolor[HTML]{FFFFFF}\textbf{\begin{tabular}[c]{@{}l@{}}Hispanic/LatinX vs.\\  Non-Hispanic/LatinX\end{tabular}}} \\ \hline \hline
\rowcolor[HTML]{FFFFFF} 

\multicolumn{1}{l|}{\cellcolor[HTML]{FFFFFF}\textbf{Quest}} & \multicolumn{1}{l|}{\cellcolor[HTML]{FFFFFF}1.0} & \multicolumn{1}{l|}{\cellcolor[HTML]{FFFFFF}1.6} & \multicolumn{1}{l}{\cellcolor[HTML]{FFFFFF}1.3} \\ \hline
\rowcolor[HTML]{FFFFFF} 

\multicolumn{1}{l|}{\cellcolor[HTML]{FFFFFF}\textbf{Point}} & \multicolumn{1}{l|}{\cellcolor[HTML]{FFFFFF}1.8} & \multicolumn{1}{l|}{\cellcolor[HTML]{FFFFFF}1.1} & \multicolumn{1}{l}{\cellcolor[HTML]{FFFFFF}1.1} \\ \hline
\rowcolor[HTML]{FFFFFF} 

\multicolumn{1}{l|}{\cellcolor[HTML]{FFFFFF}\textbf{Level}} & \multicolumn{1}{l|}{\cellcolor[HTML]{FFFFFF}0.6} & \multicolumn{1}{l|}{\cellcolor[HTML]{FFFFFF}0.8} & \multicolumn{1}{l}{\cellcolor[HTML]{FFFFFF}1.7} \\ \hline
\rowcolor[HTML]{FFFFFF} 

\multicolumn{1}{l|}{\cellcolor[HTML]{FFFFFF}\textbf{Badge}} & \multicolumn{1}{l|}{\cellcolor[HTML]{FFFFFF}0.5} & \multicolumn{1}{l|}{\cellcolor[HTML]{FFFFFF}1.1} & \multicolumn{1}{l}{\cellcolor[HTML]{FFFFFF}1.2} \\ \hline
\rowcolor[HTML]{FFFFFF} 

\multicolumn{1}{l|}{\cellcolor[HTML]{FFFFFF}\textbf{ProgressBar}} & \multicolumn{1}{l|}{\cellcolor[HTML]{FFFFFF}0.7} & \multicolumn{1}{l|}{\cellcolor[HTML]{FFFFFF}0.8} & \multicolumn{1}{l}{\cellcolor[HTML]{FFFFFF}1.0} \\ \hline
\rowcolor[HTML]{FFFFFF} 

\multicolumn{1}{l|}{\cellcolor[HTML]{FFFFFF}\textbf{Map}} & \multicolumn{1}{l|}{\cellcolor[HTML]{FFFFFF}0.7} & \multicolumn{1}{l|}{\cellcolor[HTML]{FFFFFF}0.7} & \multicolumn{1}{l}{\cellcolor[HTML]{FFFFFF}1.3} \\ \hline
\rowcolor[HTML]{FFFFFF} 

\multicolumn{1}{l|}{\cellcolor[HTML]{FFFFFF}\textbf{Stats}} & \multicolumn{1}{l|}{\cellcolor[HTML]{FFFFFF}0.7} & \multicolumn{1}{l|}{\cellcolor[HTML]{FFFFFF}1.1} & \multicolumn{1}{l}{\cellcolor[HTML]{FFFFFF}2.0} \\ \hline
\rowcolor[HTML]{FFFFFF} 

\multicolumn{1}{l|}{\cellcolor[HTML]{FFFFFF}\textbf{Choice}} & \multicolumn{1}{l|}{\cellcolor[HTML]{FFFFFF}0.8} & \multicolumn{1}{l|}{\cellcolor[HTML]{FFFFFF}1.0} & \multicolumn{1}{l}{\cellcolor[HTML]{C0C0C0}2.2*} \\ \hline
\rowcolor[HTML]{FFFFFF} 

\multicolumn{1}{l|}{\cellcolor[HTML]{FFFFFF}\textbf{Economy}} & \multicolumn{1}{l|}{\cellcolor[HTML]{FFFFFF}1.2} & \multicolumn{1}{l|}{\cellcolor[HTML]{FFFFFF}0.9} & \multicolumn{1}{l}{\cellcolor[HTML]{FFFFFF}0.9} \\ \hline
\rowcolor[HTML]{FFFFFF} 

\multicolumn{1}{l|}{\cellcolor[HTML]{FFFFFF}\textbf{TimePressure}} & \multicolumn{1}{l|}{\cellcolor[HTML]{FFFFFF}1.7} & \multicolumn{1}{l|}{\cellcolor[HTML]{FFFFFF}0.8} & \multicolumn{1}{l}{\cellcolor[HTML]{FFFFFF}1.2} \\ \hline
\rowcolor[HTML]{FFFFFF} 

\multicolumn{1}{l|}{\cellcolor[HTML]{FFFFFF}\textbf{Cooperation}} & \multicolumn{1}{l|}{\cellcolor[HTML]{FFFFFF}0.9} & \multicolumn{1}{l|}{\cellcolor[HTML]{FFFFFF}1.1} & \multicolumn{1}{l}{\cellcolor[HTML]{FFFFFF}1.2} \\ \hline
\rowcolor[HTML]{FFFFFF}

\multicolumn{1}{l|}{\cellcolor[HTML]{FFFFFF}\textbf{Renovation}} & \multicolumn{1}{l|}{\cellcolor[HTML]{FFFFFF}0.6} & \multicolumn{1}{l|}{\cellcolor[HTML]{FFFFFF}1.5} & \multicolumn{1}{l}{\cellcolor[HTML]{FFFFFF}0.8} \\ \hline
\rowcolor[HTML]{FFFFFF} 

\multicolumn{1}{l|}{\cellcolor[HTML]{FFFFFF}\textbf{Puzzle}} & \multicolumn{1}{l|}{\cellcolor[HTML]{FFFFFF}0.6} & \multicolumn{1}{l|}{\cellcolor[HTML]{FFFFFF}1.3} & \multicolumn{1}{l}{\cellcolor[HTML]{FFFFFF}1.1} \\ \hline
\rowcolor[HTML]{FFFFFF} 

\multicolumn{1}{l|}{\cellcolor[HTML]{FFFFFF}\textbf{Narrative}} & \multicolumn{1}{l|}{\cellcolor[HTML]{FFFFFF}0.5} & \multicolumn{1}{l|}{\cellcolor[HTML]{FFFFFF}1.4} & \multicolumn{1}{l}{\cellcolor[HTML]{FFFFFF}0.9} \\ \hline
\rowcolor[HTML]{FFFFFF} 

\multicolumn{1}{l|}{\cellcolor[HTML]{FFFFFF}\textbf{Storytelling}} & \multicolumn{1}{l|}{\cellcolor[HTML]{C0C0C0}0.3**} & \multicolumn{1}{l|}{\cellcolor[HTML]{FFFFFF}1.2} & \multicolumn{1}{l}{\cellcolor[HTML]{FFFFFF}0.9} \\ \hline
\rowcolor[HTML]{FFFFFF} 

\multicolumn{1}{l|}{\cellcolor[HTML]{FFFFFF}\textbf{Novelty}} & \multicolumn{1}{l|}{\cellcolor[HTML]{FFFFFF}0.8} & \multicolumn{1}{l|}{\cellcolor[HTML]{FFFFFF}1.0} & \multicolumn{1}{l}{\cellcolor[HTML]{FFFFFF}0.6} \\ \hline
\rowcolor[HTML]{FFFFFF} 

\multicolumn{1}{l|}{\cellcolor[HTML]{FFFFFF}\textbf{SocialPressure}} & \multicolumn{1}{l|}{\cellcolor[HTML]{FFFFFF}0.9} & \multicolumn{1}{l|}{\cellcolor[HTML]{FFFFFF}1.5} & \multicolumn{1}{l}{\cellcolor[HTML]{FFFFFF}1.3} \\ \hline
\rowcolor[HTML]{FFFFFF} 

\multicolumn{1}{l|}{\cellcolor[HTML]{FFFFFF}\textbf{PublicLeaderboard}} & \multicolumn{1}{l|}{\cellcolor[HTML]{FFFFFF}1.6} & \multicolumn{1}{l|}{\cellcolor[HTML]{FFFFFF}1.1} & \multicolumn{1}{l}{\cellcolor[HTML]{FFFFFF}0.9} \\ \hline
\rowcolor[HTML]{FFFFFF} 

\multicolumn{1}{l|}{\cellcolor[HTML]{FFFFFF}\textbf{PrivateRank}} & \multicolumn{1}{l|}{\cellcolor[HTML]{FFFFFF}0.9} & \multicolumn{1}{l|}{\cellcolor[HTML]{FFFFFF}0.9} & \multicolumn{1}{l}{\cellcolor[HTML]{FFFFFF}1.2} \\ \hline
\rowcolor[HTML]{FFFFFF} 

\multicolumn{1}{l|}{\cellcolor[HTML]{FFFFFF}\textbf{CompetitiveTask}} & \multicolumn{1}{l|}{\cellcolor[HTML]{FFFFFF}0.7} & \multicolumn{1}{l|}{\cellcolor[HTML]{FFFFFF}0.6} & \multicolumn{1}{l}{\cellcolor[HTML]{FFFFFF}0.7} \\ \hline \hline
\multicolumn{4}{l}{\textit{Significance codes: * \(p < 0.10\), ** \(p < 0.05\).}} \\
\multicolumn{4}{l}{\textit{\begin{tabular}[c]{@{}l@{}}Note: Odds ratio greater than 1 means the first segment has greater \\chances of reporting higher preference in the game element than the \\second. A Ratio less than 1 means the opposite. The preference was \\coded from the survey Likert questions.\end{tabular}}}
\end{tabular}
\end{table}

\subsection{Students' preferences of game elements concerning diverse characteristics}

To answer RQ2: \textit{\rqtexttwo}, we assessed the odds ratios to evaluate the preferences of each subgroup, as observed in Table~\ref{tab:oddsratio}. The preference levels were classified as high when respondents chose \textit{agree} or \textit{strongly agree} and as low when selections ranged from \textit{neutral}, \textit{disagree}, or \textit{strongly disagree}.

Our findings from the odds ratio test show no significant statistical differences in game element preferences among most groups. Only participants who align with Tim's GenderMag personality traits have a higher preference for game elements such as Storytelling (0.3) than Abi. Additionally, Hispanics/LatinX have greater odds of engagement in a gamified learning environment that offers the Choice element (2.2x) than non-Hispanics/LatinX.

We conducted the Mann-Whitney U test to analyze the differences in game element preferences across various groups, including persona, gender, and ethnicity. This nonparametric statistical test is beneficial for comparing two independent groups without assuming normal distribution in the data~\cite{feltovich2003nonparametric}. The test results for comparing game element preferences by persona yielded a p-value of 0.5937, by gender a p-value of 0.6691, and by ethnicity a p-value of 0.19. The result does not offer significant statistical differences in game element preferences among the groups.

\rqtwo[
    \tcblower
    \textbf{Answer:} No significant statistical differences were observed across persona, gender, and ethnicity in most cases. The exceptions were participants who exhibited Tim’s GenderMag personality traits, showing a preference for Storytelling, and Hispanic/LatinX individuals, who were more likely to prefer the Choice element.
]{}

\section{Discussion}
\label{sec:discussion}

This section explores the insights derived from our research findings and identifies opportunities for further exploration.

\textbf{Enhancing engagement.} Gamification has been widely implemented in educational settings and teaching methodologies to boost student engagement and motivation by integrating game elements~\cite{dichev2017gamifying}. Our findings indicate that game elements with performance-oriented features, such as Stats, Maps, Levels, Points, Progress Bars, and Badges, are essential for engaging students. They offer performance feedback and foster a sense of advancement. Furthermore, elements from the personal dimension, including Novelty, Puzzles, Quests, and Renovation, significantly contribute to engagement by presenting new challenges and content, which suggests an enhancement in learning activities. Given that motivation and engagement are central themes in gamification research~\cite{leitao2022systematic}, a gamified approach incorporating these elements is well-positioned to increase student engagement in learning the OSS contribution process.

\textbf{Diverse preferences among game elements.} Zahedi et al.~\cite{zahedi2021gamification} affirm that few studies demonstrate the differences between men and women regarding the performance impact of gamification; those that exist also show mixed results. One gender study demonstrated that gamification (points, badges, levels) improved male students' performance, but no improvement was observed for female students~\cite{pedro2015does}. In accordance with these previous studies, our findings show no significant statistical differences across persona, gender, and ethnicity for most cases, except Storytelling for Tim's persona and Choice for Hispanic/Latino. Additionally, the limited sample of empirical studies that explicitly explored gender are centered on the impacts of video games and not gamifcation~\cite{zahedi2021gamification}. Therefore, future research could explore why certain game elements appeal more to specific populations and the reasons for these preferences.

\textbf{Exploring the gamification considering self-determination theory.} Self-determination theory, established by Deci and Ryan~\cite{deci1985general}, has been used to conceptualize gamification. This theory identifies three psychological needs, including competence or the desire for mastery, reflecting the innate need to feel proficient and successful, driving people toward skill enhancement and pursuing challenges that foster growth~\cite{ryan2000intrinsic, hsu2004people}. Our findings suggest that various game elements significantly contribute to the perception of advancement (e.g., Stats, Maps, Levels, Points, Progress Bars, Badges, and Quests), raising the question of whether they are aligned with the psychological need for competence. Additionally, future studies can investigate how game elements can be strategically implemented to satisfy competence among the students, thereby harnessing a potent source of motivation for learners.

Autonomy is another essential psychological need, and it embodies the intrinsic desire for self-direction and the ability to choose paths freely and meaningfully. According to participants' open responses, our findings pinpoint the choice element as the only satisfying autonomy need. The segmented analysis indicates that participants identifying as Tim, Man, Hispanic/LatinX ranked the Choice element highly. This observation prompts further exploration into how autonomy can be facilitated within gamified environments to enrich both the learning experience and outcomes by granting learners the freedom to navigate and create their educational journey.

\textbf{Reevaluating competition in gamified learning environment.} Previous studies have claimed the double-edged nature of leaderboards and competitive elements in educational settings, noting their potential to boost competition but also to hinder overall performance, cooperation, problem-solving and even foster negative classroom dynamics, thereby increasing demotivation~\cite{leitao2022systematic}. Upward comparisons among students have been shown to potentially lead to adverse effects and diminished academic self-concept~\cite{dijkstra2008social}. Our findings echo this ambivalence towards competitive game elements such as Public Leaderboards, Private Rankings, and Competitive Tasks, which emerged as the least favored among participants (see Figure~\ref{fig:toprank}). While some literature accentuates the appeal of individual and team competition, other research points to its detrimental impact on practical activities. For example, Nevin et al.~\cite{nevin2014gamification} stated that the ability to compete individually and in teams appealed to many students. Simultaneously, Çakıroğlu et al.~\cite{akrolu2017gamifying} reported that competition came to the forefront as a negative element in practical activities. The divergent perspectives on the same game elements suggest a complex relationship between competition and motivation, highlighting the need for further investigation. 

\textbf{Gamification in OSS.} The applicability of game elements to actual OSS contributions is crucial. Concrete examples of how gamification could be integrated into OSS practices include using progress bars for tracking contributions, leaderboards for incentivizing participation, and badges for recognizing achievements. Another possibility would be to address the barriers of newcomers needing a deep understanding of the codebase by using game elements such as quests through incremental and smaller manageable contributions like bug reporting and documentation updates. Barriers that prevent contributions from newcomers~\cite{steinmacher2015social} demotivate students in OSS, especially those from diverse backgrounds, including different cognitive styles, genders, and ethnicities. A survey by Trinkenreich et al.~\cite{trinkenreich2021women} highlights existing barriers in OSS communities, including toxic culture, stereotyping, and uninclusive practices. Gamification can engage and attract a diverse group of contributors. However, it is important to recognize that while gamification can enhance engagement, it may not address deeper systemic issues impacting OSS communities. We believe that gamification can complement efforts to address these systemic issues, but it does not replace other initiatives.

\textbf{Implications.} Our research emphasizes the need for designers to carefully select and implement game elements that resonate with diverse student preferences to enhance engagement. We showed that Performance-related and Personal elements are well-accepted by participants from different backgrounds (crosscutting gender, ethnicity, and cognitive styles). Therefore, they may impact the engagement of more students in a more equal and inclusive environment. Moreover, educators can use our findings to gamify the learning experience, the OSS contribution process, and beyond. Understanding that students have diverse preferences for game elements, educators can tailor their teaching strategies and incorporate a mix of performance and personal dimension elements to meet these preferences. Researchers can also replicate our study and further explore our data, which is available in our replication package. 

\section{Limitations}
\label{sec:threatstovalidity}

\textbf{Sampling bias.} As detailed in Section~\ref{sec:methodology}, we used Prolific to recruit participants, and we attempted to balance our sampling with the following criteria: (i) Men/LatinX; (ii) Non-Men/LatinX; (iii) Men/Non-LatinX; and (iv) Non-Men/Non-LatinX. However, we did not receive enough responses from non-men. Moreover, we recognize that our sample may carry biases that are not immediately apparent. Therefore, our conclusions are only valid for our participants. Future research should aim for a more balanced and globally representative sample to enhance the universality of the findings.
Additionally, we segmented the dataset to examine students' preferences behind certain game elements. The process involved dividing the dataset into smaller, more manageable groups for focused analysis. While segmenting the data can sometimes present challenges for statistical analysis due to reduced group sizes, each group contained a minimum of 40 responses. The volume of data is adequate for conducting \textit{Chi-Square Tests}~\cite{sureiman2013conceptual}, the statistical significance test we employed in our odds ratio analysis for RQ2. 

\textbf{Response biases.} We introduced a preliminary context-setting stage to mitigate the impact of such biases. The context was designed to provide a scenario for participants, encouraging them to anchor their responses within a more objective framework to capture more accurate and meaningful reflections of their experiences.

\textbf{Construct validity.} To enhance construct validity, we based our survey on the game elements taxonomy proposed by Toda et al.~\cite{toda2019analysing}. Moreover, we employed pilot studies to test and collect feedback about our instrument to mitigate this threat. The iterative studies enabled us to refine and validate our survey tool, ensuring its effectiveness and reliability in capturing the intended constructs. Another concern in our study is the participants' prior familiarity with the game elements discussed in the survey. To address this issue, we provided a contextualized explanation of each game element, ensuring a more informed response process. Regarding the hypothetical nature of the survey and the potential gap between these scenarios and real-life OSS contributions, we recognize this paper is an initial step toward developing a gamified learning environment to support students' learning about the OSS contribution process. The survey helps to understand perceived value and user receptiveness, which are relevant from the user-centered design perspective.

\textbf{Subjectivity.} We employed qualitative procedures to classify the open questions responses. These procedures are subject to subjectivity/interpretation bias. To mitigate this potential bias, we adopted a multi-faceted approach involving the collaboration of multiple researchers. The team engaged in continuous comparative analysis and reached conclusions through a process of negotiated agreement. Each team member has extensive experience in qualitative methods and OSS.

\section{Conclusion}
\label{sec:conclusion}

We surveyed 115 participants, asking which game elements would boost engagement in a gamified learning environment for teaching the OSS contribution process. Enhancing student engagement has the potential to improve the contribution experience and expand the contributor base, making OSS projects more accessible and appealing to a diverse range of contributors. 

Our findings highlight the importance of integrating game elements that offer students constructive feedback on their performance and foster a sense of progression and achievement. In future work, we aim to design and develop a gamified environment with such game elements.

\section*{Acknowledgments}
The National Science Foundation (NSF) partially supports this work under grant numbers 2236198, 2247929, 2303042, and 2303612.  Katia Romero Felizardo is funded by a research grant from the Brazilian National Council for Scientific and Technological Development (CNPq), Grant $302339/2022-1$. 


\bibliographystyle{IEEEtran}
\bibliography{IEEEabrv,bibtex.bib}

\end{document}